\documentclass[12pt]{article}
\usepackage{epsfig}
\usepackage{xcolor}

\usepackage{graphicx}
\usepackage{cancel}
\usepackage{amssymb,amsmath}

\def\harr#1#2{\smash{\mathop{\hbox to .3in{\rightarrowfill}}
 \limits^{\scriptstyle#1}_{\scriptstyle#2}}}

\def\s2{\frac{1}{\sqrt2}}

\def\beqa{\begin{eqnarray}}
\def\eeqa{\end{eqnarray}}

\def\Dsl{\,\raise.15ex\hbox{/}\mkern-13.5mu D} 
\def\d3{d^3}

\newcommand{\be}{\begin{equation}}
\newcommand{\ee}{\end{equation}}
\newcommand{\beq}{\begin{eqnarray}}
\newcommand{\eeq}{\end{eqnarray}}

\usepackage{amssymb,amsmath}

\usepackage{mathrsfs}

\def\be{\begin{equation}}
\def\ee{\end{equation}}
\def\beqa{\begin{eqnarray}}
\def\eeqa{\end{eqnarray}}

\begin{document}

\begin{center}
\Large{\bf Emergent Dark Matter and Dark Energy from a Lattice
Model} \vspace{0.5cm}

\large Luis Lozano\footnote{e-mail address:
{\tt lalozanom@tec.mx}}, Hugo Garc\'ia-Compe\'an\footnote{e-mail address: {\tt
compean@fis.cinvestav.mx}}

{\small \em Departamento de F\'{\i}sica, Centro de
Investigaci\'on y de Estudios Avanzados del IPN,}\\
{\small\em P.O.\ Box 14-740, CP.\ 07000, Ciudad de M\'exico, M\'exico}
\vspace*{0.5cm}

\vspace*{1.5cm}
\end{center}

\begin{abstract}

We propose a quantum bosonic qubit model on a fcc lattice that
realizes the canonical source structure of mimetic dark matter as
a defect of a rank-two lattice Gauss law. The standard
contribution from general relativity is implemented similarly to
previous works in the literature, while the mimetic sector
modifies the constraint equations through additional source
terms. Different theories such as mimetic dark matter, vector
mimetic dark matter, and tensor-vector-scalar models are
implemented on the lattice. In all these cases, a generalized
Gauss law incorporates an additional Gauss-law (topological)
defect depending on the type of generalization, but always
fitting into the structure of the defects from the general
relativity contribution. We also derive the resulting
charge-selection rules for the defect sectors and show, in a
worked example, that a localized defect sources a long-range
rank-two field whose trace-free representative is exactly the
Bowen--York momentum solution of canonical gravity. The
mimetic constraint is treated in its
full ADM form, retaining the normal derivative of the scalar
field, and the known ghost and gradient instabilities of the
minimal continuum mimetic theory are summarized. The lattice
construction is therefore presented as a formal realization of
the canonical source structure rather than as a complete
cosmological model.

\vskip 1truecm

\end{abstract}

\bigskip

\newpage

\section{Introduction}
\label{sec:intro}

Diversity of the laws describing physical phenomena is one of the
more intriguing mysteries in physics. Through the years the emergence
principle has been extremely useful in condensed matter  physics
(CMP) (see for instance,
\cite{Anderson:1972pca,bookWen,Levin:2004js,Wen:2017usd}). This
principle has permeated in other areas of physics. For instance, in
theoretical high energy physics, and recently, in string theory.

At the present time, it is believed that spacetime described by the
general theory of relativity (GR) is a long-distance (or low
energies) approximation of an underlying theory at the Planck scale
with fundamental degrees of freedom of different nature beyond that the
one of the usual metric. Thus, GR is regarded as an effective theory valid at
distances much more greater than the Planck length $L_P$ (for a
review, see for instance, \cite{Carlip:2012wa}).

In the case of string theory, the emergence principle is naturally
incorporated in the AdS/CFT correspondence \cite{Aharony:1999ti}. It
is well known that spacetime is a derived object from local degrees
of freedom described by a supersymmetric gauge theory in the
boundary of that spacetime.

Recently, in CMP developments regarding topological phases of matter
have motivated new systematic implementations of theories where the
gravitational interactions and the spacetime itself is emergent.
Some of these examples were given in Refs.
\cite{Gu:2009jh,Xu:2006faa}. In these references a graviton model is
obtained in GR and in the context of Horava-Lifshitz theory of
gravity \cite{Xu:2010eg}. A more recent attempt to obtain the
fundamental string from a lattice model was given in
\cite{Lozano:2019oxi}.

On the other hand, a new theory incorporating dark matter in general
relativity was proposed in \cite{Chamseddine:2013kea}. This theory
was called {\it Mimetic Dark Matter} (see \cite{Sebastiani:2016ras}
for a review). In this theory, the usual dynamical degrees of
freedom of GR, $g_{\mu \nu}$, split out into an auxiliary metric and
a scalar field. In this reference, it was argued that the scalar
field component behaves as dark matter. In subsequent works the {\it
dark energy} behavior was also incorporated. Since the development
of Mimetic Dark Matter in \cite{Chamseddine:2013kea}, and the
extension to Vector Field Mimetic Dark Matter, TeVeS, and
Inhomogeneous Mimetic Dark Energy, and Mimetic Gravity
\cite{Chaichian:2014qba, Chamseddine:2016uyr, Kluson:2017iem,
Cognola:2016gjy}, there have been applications found in cosmology
\cite{Chamseddine:2014vna, Chamseddine:2016uef, Gorji:2018okn,
Bezerra:2019bgt, Myrzakulov:2015qaa, Dutta:2017fjw,
Casalino:2018tcd, Casalino:2018wnc}, in Einstein's gravity and
geometry \cite{Barvinsky:2013mea, Khalifeh:2019zfi,
Vagnozzi:2017ilo}, and in problems of singularities and black holes
\cite{Chamseddine:2016ktu, Chamseddine:2019pux}. Mimetic
model-building has also been carried out for modified gravities and
anisotropic backgrounds, including mimetic $f(R)$ and $f(R,T)$
cosmologies obtained through Noether-symmetry methods
\cite{Momeni:2015mimetic} and mimetic matter in LRS Bianchi type-I
models \cite{GudekliDemir:2021}. Very recently, the
production of mimetic dark matter directly from inflation has also
been analyzed \cite{Chamseddine:2026fti}.

In the present work we propose a lattice model for mimetic gravity.
In this model, mimetic dark-sector source terms are represented as
Gauss-law defects in a qubit-inspired fcc lattice construction.
In particular, we focus on how the canonical mimetic source structure is realized on the fcc rank-two lattice graviton model of Xu and Ho\v{r}ava \cite{Xu:2010eg,Xu:2006faa,Gu:2009jh}; the cosmological viability of any specific mimetic completion is outside the scope of the present formal construction.
Our interest is to keep developing the emergence of fields from lattice models
and provide a description of this gravitational phenomenon in terms of qubit defects in the lattice \cite{Wen:2017usd, Fradkin:1978th}. The present lattice construction is canonical and quasi-static: it is formulated on a fixed spatial slice, while the normal derivatives of the continuum fields are retained through the corresponding canonical momenta. Thus it is not our aim to give a description of time-dependent cosmological phenomena of dark matter and dark energy themselves.

We situate this construction within two recent threads in the
literature. On the gravity-theory side, the mimetic framework has
been substantially refined since 2015 through Hamiltonian and
effective-field-theory analyses
\cite{Arroja:2015wpa,Ganz:2019pkl,Langlois:2018jdg,deCesare:2020swb},
including its identification with a subclass of degenerate
higher-order scalar-tensor (DHOST) theories \cite{Langlois:2018jdg}
and its disformal-symmetry-based formulation
\cite{Domenech:2023ryc}. On the condensed-matter side, the fcc
rank-two construction belongs to the broader family of higher-rank
gauge theories that emerge from constrained bosonic models on the
lattice, and whose connection with fracton physics has been
clarified in recent reviews
\cite{Pretko:2017fbf,Pretko:2017xar,You:2024prv}. The present
lattice realization sits at the intersection of these threads.

The article is organized as follows. In Sec. \ref{sec:prelim} we give the basic
facts about mimetic gravity, which will be important in subsequent
sections. In Sec. \ref{sec:grat} we give the implementation in the lattice of
different models of dark matter and dark energy. The mimetic dark
matter models with scalars and vector fields is also implemented.
The inhomogeneous dark energy model and the mimetic
tensor-vector-scalar gravity is also described. Finally, in Sec.
\ref{sec:final} a summary of our results and some final remarks are given.

\section{Preliminaries on Mimetic Dark Matter}
\label{sec:prelim}

In this section we briefly overview some basics facts on mimetic
dark matter \cite{Chamseddine:2013kea, Chaichian:2014qba,
Kluson:2017iem, Barvinsky:2013mea, Ganz:2018mqi}, which will be
necessary in the following sections. We are interested in giving
some notation and conventions for future reference. Thus in this
section we give a brief overview in order to present the all the
material of mimetic dark matter in a concise way. However a few of
the equations presented here, the Hamiltonians, will be repeated for
convenience in subsequent sections.  We follow mainly the references
\cite{Chaichian:2014qba, Kluson:2017iem, Ganz:2018mqi,
Arnowitt:1962hi, Gourgoulhon:2007ue}.

We start by writing down the physical metric $g^{\textrm{phys}}_{\mu
\nu}$ in terms of the fundamental metric $g_{\mu \nu}$ and the
gradient of a scalar field $\phi$. This is given by \be
g^{\textrm{phys}}_{\mu \nu} = \bigg(-g^{\alpha \beta}
\partial_\alpha \phi \partial_\beta \phi \bigg) g_{\mu \nu}.\ee

The physical metric is Weyl invariant. Thus, mimetic dark matter is
a conformal extension of Einstein theory of gravitation. We will
take \be g^{\textrm{phys}}_{\mu \nu} = \Phi^2 g_{\mu \nu}
\label{ONE}, \ee where we will be working with the gauge fixing:
$\Phi - 1 = 0.$

The condition $\Phi = 1$ should be understood as a Weyl gauge
fixing rather than as the physical mimetic constraint itself.
The covariant mimetic constraint, obtained by substituting
Eq.~(\ref{ONE}) into the original parametrization
$g^{\textrm{phys}}_{\mu\nu}=(-g^{\alpha\beta}\partial_\alpha\phi
\partial_\beta\phi)\,g_{\mu\nu}$, reads
\be
g_{\rm phys}^{\mu\nu}\,\partial_\mu\phi\,\partial_\nu\phi=-1,
\label{eq:covmimetic}
\ee
and it is this four-dimensional condition that fixes the canonical
source structure of the scalar sector
\cite{Chamseddine:2013kea,Arroja:2015wpa,Ganz:2018mqi}. In the ADM
decomposition introduced below, the future-directed timelike unit
normal $n^\mu$ satisfies $\nabla_n\phi\equiv
n^\mu\partial_\mu\phi=(1/N)(\partial_t\phi-N^i\partial_i\phi)$,
so that, using $g^{\mu\nu} = -n^\mu n^\nu + h^{\mu\nu}$,
\be
g^{\mu\nu}\partial_\mu\phi\,\partial_\nu\phi
= -(\nabla_n\phi)^2
+ h^{ij}\partial_i\phi\,\partial_j\phi,
\label{eq:ADMdecomp}
\ee
and Eq.~(\ref{eq:covmimetic}) is equivalent to
\be
(\nabla_n\phi)^2 = 1 + h^{ij}\partial_i\phi\,\partial_j\phi.
\label{eq:ADMmimetic}
\ee
The normal derivative of $\phi$ therefore enters the mimetic
constraint on equal footing with its spatial gradients; in
particular, Eq.~(\ref{eq:ADMmimetic}) cannot be reduced to a purely
spatial relation without an explicit further gauge choice. Of
course, in the cosmological unitary or dust-time gauge $\phi=t$
\cite{deCesare:2020swb} one has $\partial_i\phi=0$ and the
constraint reduces to $(\nabla_n\phi)^2=1$, but for inhomogeneous
configurations the full Eq.~(\ref{eq:ADMmimetic}) must be used.

Before writing the Hamiltonian, we will need to use the ADM
decomposition of $g_{\mu \nu}$ (see \cite{Chaichian:2014qba,
Kluson:2017iem, Ganz:2018mqi, Arnowitt:1962hi, Gourgoulhon:2007ue}):
$$ g_{00} = -N^2 + N_i h^{ij} N_j, \qquad g_{0i} = N_i, \qquad
g_{ij} = h_{ij}, $$ \be g^{00} = - {1 \over N^2}, \qquad g^{0i} =
{N^{i} \over N^2}, \qquad g^{ij} = h^{ij} - {N^{i} N^{j} \over N^2},
\label{TWO} \ee where $h^{ij}$ is the inverse to the induced metric
$h_{ij}$ on the constant time surfaces,  $N$ and $N^{i}$ are the
lapse and shift functions from the ADM decomposition, and $N^{i} =
h^{ij} N_j. $

We will be working with a Hamiltonian of the form (see
\cite{Chaichian:2014qba, Kluson:2017iem, Ganz:2018mqi,
Malaeb:2014vua}) \be H=\int d^3 x (N \mathcal{H}_T + N^{i}
\mathcal{H}_{i}), \label{THREE} \ee with \be \mathcal{H}_T =
\mathcal{H}_T^{GR} + \mathcal{H}_T^{\circ} \approx 0. \label{FOUR}
\ee where \be \mathcal{H}^{GR}_T = {2 \over \sqrt{h}} \pi^{ij} {\cal
G}_{ijkl} \pi^{kl} - {1 \over 2} \sqrt{h} R, \ee where $\pi^{ij}$ is
the conjugate momentum to $h_{ij}$ given by $\pi^{ij} = {1\over 2}
\sqrt{h} {\cal G}^{ijkl} K_{kl}$, ${\cal G}_{ijkl} = {1 \over 2}
(h_{ik} h_{jl} + h_{il} h_{jk}) -{1\over 2} h_{ij} h_{kl}$ is the
DeWitt metric, ${\cal G}^{ijkl} = {1 \over 2} (h^{ik} h^{jl} +
h^{il} h^{jk}) - h^{ij} h^{kl}$ is its inverse metric and $h$ is the
determinant of $h_{ij}$. $K_{ij}$ is the extrinsic curvature and it
is given by $K_{ij} = {1 \over 2N} ( {\partial h_{ij} \over
\partial t} -D_iN_j- D_jN_i)$, where $D_i$ is the covariant derivative with respect
to $h_{ij}$. It is the standard term from general relativity, which
will be obtained in Section \ref{sec:grat}, and
$\mathcal{H}^{\circ}_T $ is the term related to the fields that will
be added on the following sections. For instance, for only mimetic
dark matter we have \be \mathcal{H}^{\circ}_T = \mathcal{H}^{\phi}_T
= {1\over 2 \lambda \sqrt{h}} p_\phi^2 +{1 \over 2} \sqrt{h}
\lambda(1 + h^{ij}\partial_i\phi \partial_j\phi), \label{onebb}
 \ee where $p_\phi$ is the conjugate momentum with respect to $\phi$ given by
 $p_\phi = \sqrt{h} \lambda \nabla_n \phi$, $\nabla_n \phi ={1 \over N}(\partial_t \phi
-N^i \partial_i \phi)$ and $\lambda$ is a Lagrange multiplier whose
variation, combined with the canonical relation
$p_\phi=\sqrt{h}\,\lambda\,\nabla_n\phi$, enforces the full ADM
mimetic constraint Eq.~(\ref{eq:ADMmimetic}) on the constraint
surface. Its canonical momentum $p_\lambda$ is given by $p_\lambda \approx 0$.
The vector constraint is given by \be {\cal H}_i = p_\phi
\partial_i \phi - 2 h_{ij} D_k\pi^{jk}. \ee

For the vector field model of mimetic dark matter
$\mathcal{H}^{\circ}_T$ is given by
$$
\mathcal{H}^{\circ}_T  = \mathcal{H}^{u}_T =  {1\over 2
\mu^2\sqrt{h}} p^i h_{ij} p^j - {1 \over 2} \sqrt{h} \lambda \bigg(1
+ u_i h^{ij}u_j - u_{\bf n}^2\bigg)
$$
\be + {\mu^2 \over 4} \sqrt{h} h^{ik} h^{jl} (D_iu_j -
D_ju_i)(D_ku_l - D_lu_k) - u_{\bf n} D_i p^i, \ee where $\mu^2$ is a
parameter with the dimensions of mass squared and $u_{\bf n}$ is the
normal component of the vector field $u_i$ with respect the constant
time hypersurfaces in the ADM decomposition.

The vector constraint is \be {\cal H}_i = \partial_i u_j p^j -
\partial_j(u_i p^j) - 2 h_{ik}D_j \pi^{kj}, \label{twobb}\ee where
$p^i$ is the conjugate momentum to the vector variable $u_i$.

Finally, for the mimetic tensor-vector-scalar gravity the
Hamiltonian is given by
$$
\mathcal{H}^{\circ}_T  =  \mathcal{H}^{u,\phi}_T = {p_\phi^2 \over
2 \sqrt{h}} + {1\over 2} \sqrt{h} h^{ij} \partial_i \phi
\partial_j \phi + V(\phi) +  {1\over 2 \mu^2\sqrt{h}} p^i h_{ij} p^j
$$
\be -{1 \over 2} \sqrt{h} \lambda (1 + f(\phi) u_i h^{ij}u_j -
f(\phi) u_{\bf n}^2) + {\mu^2 \over 4} \sqrt{h} h^{ik} h^{jl}
(D_iu_j - D_ju_i)(D_ku_l - D_lu_k) - u_{\bf n} D_i p^i, \ee
while the vector constraint
is given by \be {\cal H}_i = p_\phi
\partial_i \phi  + \partial_i u_j p^j - \partial_j(u_i p^j) - 2 h_{ij}
D_k\pi^{jk}. \label{threebb} \ee It is important to remark that in
our procedure all constraints will be implemented on the reduced
phase space i.e. $\mathcal{H}_T \approx 0$, and $\mathcal{H}_i
\approx 0$. The term $\mathcal{H}_i $ is related to constraints
depending on the added fields. There are other terms in the
Hamiltonian in equation (\ref{THREE}) (see
\cite{Chamseddine:2013kea, Chaichian:2014qba, Ganz:2018mqi,
Golovnev:2013jxa}), but are not relevant for our purposes.

\section{Emergence of Mimetic Dark Matter and Dark Energy from Bosonic qubit
Models}
\label{sec:grat}

In the following subsections, we will introduce different models of
mimetic dark matter, and inhomogeneous dark energy  in order to give
a physical meaning to the topological defects found on equation
(\ref{THIRTEEN}) as a violation of equation (\ref{TEN}), see
\cite{Chamseddine:2013kea, Chaichian:2014qba, Chamseddine:2016uyr,
Kluson:2017iem, Barvinsky:2013mea, Malaeb:2014vua, Golovnev:2013jxa,
Chaichian:2014dfa}.

\subsection{Physical content of the construction}
\label{subsec:physcontent}

Before entering the technical construction, it is important to
point out what is claimed in this work and what is not. The
novelty of this work is not a cosmological solution, an equation
of state, or an observational prediction. The novelty is the
identification of the canonical mimetic source terms (for the
scalar, two-scalar, vector, and tensor-vector-scalar sectors) with
charged sectors of the rank-two lattice Gauss law of the fcc
bosonic model, based on the full ADM form of the mimetic
constraint derived in Sec.~\ref{sec:prelim}. This identification
has three concrete consequences.

First, it provides a criterion: a mimetic dark-sector
inhomogeneity necessarily lies in a charged (defect) sector of the
constrained Hilbert space, that is, it violates the vacuum Gauss
law (\ref{TEN}). Conversely, a charged lattice configuration
represents a scalar mimetic source only when its charge admits the
canonical form of Eq.~(\ref{NINETEEN}) and satisfies the
associated mimetic constraints. Second, the identification implies
selection rules, derived in Sec.~\ref{subsec:mimedarkmatt},
restricting which defect configurations are admissible at all on a
closed lattice. Third, it yields a quantitative long-distance
statement: the trace-free minimal-energy field profile sourced by
a localized defect reproduces, at long wavelength, exactly the
Bowen--York momentum-constraint field of a point source in
canonical gravity (Sec.~\ref{subsec:pointdefect}).

It is important to keep in mind two distinctions. The defects
considered here are the lattice representatives of the
\emph{inhomogeneous} mimetic momentum-density source; in this
precise canonical sense they realize the mimetic dark-sector source
on the lattice. We do not identify an individual lattice defect
with a phenomenological dark-matter particle or a dark-energy fluid
element: the identification is at the level of the canonical source
entering the gravitational constraint. Likewise, the homogeneous
cosmological mimetic dust sector (described in dust-time gauge
$\phi = t$, where the spatial gradients vanish) is distinct from
the inhomogeneous Gauss-law defect sector studied here; the former
contributes through the Hamiltonian constraint, the latter through
the momentum constraint. This mode of identification, in which a
long-distance sector of a microscopic lattice model is matched to a
continuum field-theoretic structure through its constraint algebra
rather than through a full dynamical equivalence, is the standard
one in the emergent-gravity literature
\cite{Gu:2009jh,Xu:2006faa,Xu:2010eg}.

It is also useful to fix the sense in which the word
\emph{emergent} is used in this paper, including in its title. We
use it in the condensed-matter sense of
Refs.~\cite{Anderson:1972pca,bookWen,Wen:2017usd}: a structure is
emergent when it arises as the long-distance, collective
description of a microscopic model whose elementary degrees of
freedom are of a different nature. In this sense the mimetic
dark-sector sources studied here are emergent: they are realized
as charged sectors of the low-energy rank-two gauge structure of a
bosonic qubit model, in the same way in which the graviton itself
is emergent in Refs.~\cite{Gu:2009jh,Xu:2006faa,Xu:2010eg}.
Emergence in this sense is a statement about the constraint-level
description of the lattice model; it does not by itself imply the
cosmological viability of any specific mimetic completion, which
is a separate question addressed in
Sec.~\ref{sec:stability-scope}.

\subsection{Standard general relativity contribution}

\begin{figure}
\begin{center}
\includegraphics[scale=0.6]{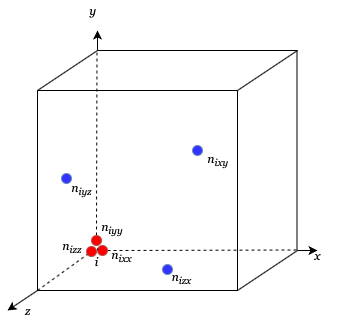}
\caption[Short caption for figure 1]{\label{Figure1} {\scriptsize A
representation of the boson numbers $\widehat{n}_{i \alpha \beta}$
defined on the site $i$, for $\alpha = \beta$ (in red); and faces $i
+ {\widehat{x} \over 2} + {\widehat{y} \over 2}$, $i + {\widehat{y}
\over 2} + {\widehat{z} \over 2}$, and $i + {\widehat{z} \over 2} +
{\widehat{x} \over 2}$, for $\alpha \neq \beta$ (in blue).
Physically, these bosonic occupation numbers are the microscopic
variables whose deviations from the mean define the rank-two field
$\pi_{i,\alpha\beta}$, the lattice counterpart of the ADM momentum
$\pi^{ij}$ (Table~\ref{tab:dictionary}); the site/face placement
implements the six components of a symmetric tensor on the
bipartite fcc structure.}}
\end{center}
\end{figure}

Now, we will use the model given in Refs.
\cite{Xu:2010eg,Xu:2006faa} (with the notation of reference
\cite{Lozano:2019oxi}) to implement the general relativity contribution of
equation (\ref{FOUR}). For this model we will go straightforward to
a bipartite lattice, for which we will use a fcc system with the
gravitational variables defined on the sites and the center of the
plaquettes (faces). In this way, we introduce the boson numbers
$\widehat{n}_{i, \alpha \beta}$ defined on each plaquette $i +
{\alpha \over 2} + {\beta \over 2}$ when $\alpha \neq \beta$, and on
each site $i=(i_x, i_y, i_z)$ when $\alpha = \beta$, as shown on the
Figure \ref{Figure1} ($i$ denotes the site of the lattice, while
$\alpha$, $\beta$ denote the directions $x$, $y$, $z$). We define
similarly the phase angle variables $\widehat{\theta}_{i, \alpha
\beta}$ conjugate to the boson numbers, and related to the boson
creation operators by $\widehat{b}_{i, \alpha \beta} \propto e^{- i
\widehat{\theta}_{i, \alpha \beta} }$, and by the commutation
relations \be [\widehat{n}_{i, \alpha \beta}, \widehat{\theta}_{i,
\gamma \delta}] = i \delta_{\alpha \gamma} \delta_{\beta \delta}.
\label{FIVE} \ee

The Hamiltonian of the system is written as follows: \be H^{GR} =
H_1 + H_2 + H_3, \label{SIX} \ee with \be H_1 = -t \sum_{\langle i,
\alpha \beta; j, \gamma \delta \rangle} \widehat{b}^{\dagger}_{i,
\alpha \beta}  \widehat{b}_{j, \gamma \delta} + {\rm h.c.},
\label{SEVEN} \ee \be H_2 = U \sum_{i, \alpha \beta}(n_{i, \alpha
\beta} - \bar{n})^2 , \label{EIGHT} \ee \be H_3 |_{(i, x)} = V(n_{i,
z x} + n_{i-z, z x} + n_{i, x y} + n_{i-y, x y} + 2n_{i, x x} +
2n_{i+x, x x} - 8)^2. \label{NINE} \ee

\begin{figure}
\begin{center}
\includegraphics[scale=0.6]{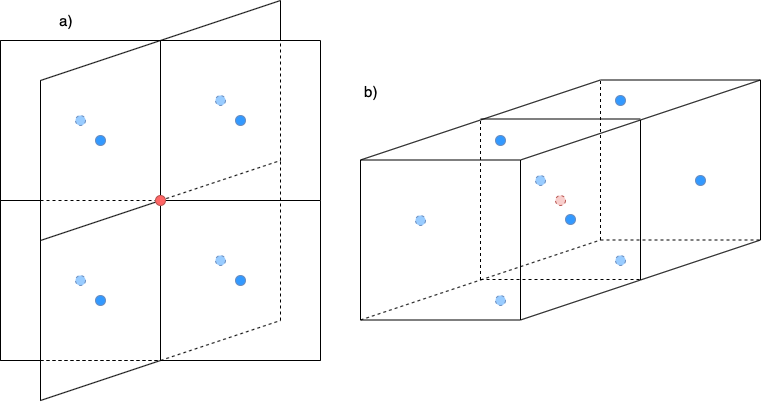}
\caption[Short caption for figure 2]{\label{Figure2} {\scriptsize The interaction of a) a site boson (in red) with its nearest neighbor plaquette bosons (in blue), and b) a plaquette boson (in red) with its nearest neighbor plaquettes (in blue) from equation (\ref{SEVEN}). Physically, these hopping processes generate, at eighth order in perturbation theory, the cosine terms of the effective Hamiltonian (\ref{ELEVEN}), whose arguments discretize the components of the linearized curvature tensor (\ref{TWELVE}); they are the microscopic origin of the emergent rank-two dynamics.}}
\end{center}
\end{figure}

The first term, $H_1$, is the nearest-neighbor hopping term between
the boson numbers, for which the sum has the brackets "$\langle$"
and "$\rangle$" indicating that only nearest site-plaquette, and
nearest plaquette-plaquette are allowed, which have the same
distance. This interactions are shown in Figure \ref{Figure2}. The
second term, $H_2$, is the on-site interaction in which $\bar{n}$
means the average boson number, which we will fix to $1$ for
simplicity, and the sum is over all the sites and boson numbers. The
quantities $n_{i, \alpha \beta}$ are the eigenvalues of the
operators $\widehat{n}_{i, \alpha \beta}$.

\begin{figure}
\begin{center}
\includegraphics[scale=0.6]{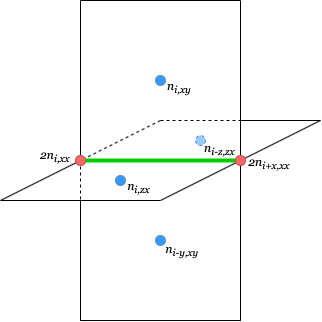}
\caption[Short caption for figure 3]{\label{Figure3} {\scriptsize
The link interaction of equation (\ref{NINE}) showing the link
$(i,x)$ (in green), the plaquette bosons (in blue), and the site
bosons (in red). Physically, this term is the energetic
enforcement of the vacuum Gauss law (\ref{TEN}): when it dominates,
configurations violating the constraint are gapped, and a mimetic
source manifests itself as a localized violation of this
constraint (a Gauss-law defect) supported on the indicated
link geometry.}}
\end{center}
\end{figure}

The last term, $H_3$, is an interaction term that only involves
links. We have explicitly shown only the term for the link $(i, x)$,
which is the link between sites $i$ and $i + x$, and is shown also
in Figure \ref{Figure3}. The other links can be written similarly.

Following \cite{Xu:2006faa} and define a symmetric second-rank
tensor $\pi_{i, \alpha \beta} = \eta_i (n_{i, \alpha \beta} -
\bar{n})$, where $\eta_i= \pm 1$, depending on the sublattice being
used, and $\bar{n} = 1$. We will take as convention $\eta_i = 1$ if
$i_x + i_y + i_z$ is even, and $\eta_i = -1$ if $i_x + i_y + i_z$ is
odd. When the term of equation (\ref{NINE}) is the dominant term, it
can be interpreted as a Gauss-like constraint \be \nabla_{\alpha}
\pi_{i, \alpha \beta} = 0, \label{TEN} \ee where a sum over repeated
indices is understood, and $\nabla_x$ is the lattice derivative in
the $x$ direction $\nabla_{x} \pi_{i, x \beta} = \pi_{i+x, x \beta}
- \pi_{i, x \beta}$. This constraint requires to keep the low energy
Hamiltonian invariant under the gauge transformation $h_{i, \alpha
\beta} \to h_{i, \alpha \beta} + \nabla_{\alpha} f_{i, \beta} +
\nabla_{\beta} f_{i, \alpha}$, where $f_{i, \alpha}$ is a vector
defined on the link $(i, \alpha)$, and $h_{i, \alpha \beta}$ the
conjugate of $\pi_{i, \alpha \beta}$.

The effective low energy Hamiltonian is given by (see
\cite{Xu:2006faa}) $$ H_{eff} = -t'_1 \sum_{i, \alpha \neq \beta}
\cos(R_{i, \alpha \beta \alpha \beta}) -t'_2 \sum_{i, \alpha \neq
\beta \neq \gamma \neq \alpha} \cos(R_{i, \alpha \beta \alpha
\gamma}) $$ \be + {1 \over K_1} \sum_{\alpha} \pi_{i, \alpha
\alpha}^2 + {1 \over K_2} \sum_{\alpha \neq \beta}\pi_{i, \alpha
\beta}^2, \label{ELEVEN} \ee where the cosine terms are generated
from the hopping terms at eight order perturbation $t'_1$, $t'_2
\sim t^8 / V^7$; and the curvature tensor is given by $$2R_{xyxy} =
\varepsilon_{zab} \varepsilon_{zcd} \nabla_a \nabla_c h_{bd}, \qquad
2R_{xzxz} = \varepsilon_{yab} \varepsilon_{ycd} \nabla_a \nabla_c
h_{bd},
$$ $$2R_{yzyz} = \varepsilon_{xab} \varepsilon_{zcd} \nabla_a \nabla_c
h_{bd}, \qquad 2R_{xyxz} = \varepsilon_{yab} \varepsilon_{zcd}
\nabla_a \nabla_c h_{bd}, $$ \be 2R_{yxyz} = \varepsilon_{xab}
\varepsilon_{zcd} \nabla_a \nabla_c h_{bd}, \qquad 2R_{zxzy} =
\varepsilon_{xab} \varepsilon_{ycd} \nabla_a \nabla_c h_{bd},
\label{TWELVE} \ee for every site.

We will not work with a dual lattice for the matters of our problem,
but we will focus on the constraint of equations (\ref{NINE}),
(\ref{TEN}). If there is a violation of this constraint, as has been
noted in references
\cite{Xu:2010eg,Xu:2006faa,Lozano:2019oxi,Gu:2009jh}, there is a
gauge charge field with phase angle $\theta_{\alpha}$ defined on the
links, coupled to the field as: \be H_c = - t \sum_{\alpha, \beta}
\cos(\nabla_{\alpha} \theta_{\beta} + \nabla_{\beta} \theta_{\alpha}
- h_{\alpha \beta}).  \label{THIRTEEN} \ee Such phase angles are
topological defects and, in the precise constraint-level sense
of Sec.~\ref{subsec:physcontent}, they can be understood as the
lattice representatives of mimetic dark-sector sources (mimetic
dark matter or inhomogeneous dark energy) in the gravitational
theory.

\subsection{Mimetic Dark Matter Model}
\label{subsec:mimedarkmatt}

Table~\ref{tab:dictionary} collects, for the reader's convenience,
the correspondence between the continuum ADM objects of
Sec.~\ref{sec:prelim} and the lattice objects of the preceding
subsection that will be used from here on.

\begin{table}
\begin{center}
\begin{tabular}{c|c|l}
\hline
Continuum ADM object & Lattice object & Role \\
\hline
$h_{ij}$ & $h_{i,\alpha\beta}$ & rank-two metric-like field \\
$\pi^{ij}$ & $\pi_{i,\alpha\beta}$ & conjugate momentum / electric tensor \\
$D_k \pi^{jk}$ & $\nabla_\alpha \pi_{i,\beta\alpha}$ & momentum constraint / Gauss law \\
$\phi(x)$ & $\phi_i$ & mimetic scalar \\
$p_\phi(x)$ & $p_{\phi_i}$ & mimetic scalar momentum \\
$\partial_i \phi$ & $\nabla_\alpha \phi_i$ & spatial gradient (links) \\
$J^{j}_{\rm mim}$ & $J^{\rm mim}_{i,\beta}$ & Gauss-law defect source \\
\hline
\end{tabular}
\caption{\label{tab:dictionary} {\scriptsize Dictionary between
continuum ADM variables and fcc lattice variables. The first three
rows are the gravitational sector of
Refs.~\cite{Xu:2006faa,Xu:2010eg}; the last four rows are the
mimetic sector constructed in this paper. The same pattern applies
to the two-scalar, vector, and tensor-vector-scalar sources of
Secs.~\ref{subsec:inhodarkener}--\ref{subsec:mimeteves}.}}
\end{center}
\end{table}

For the mimetic dark matter model, we will follow the line given in
\cite{Chaichian:2014qba, Kluson:2017iem}, write the physical metric
$g^{\textrm{phys}}_{\mu \nu}$ in terms of the fundamental metric
$g_{\mu \nu},$ and a scalar field $\phi$ that we will add.
Substituting this in the equation (\ref{ONE}), and with the gauge
fixing $\Phi - 1 = 0$ we obtain that \be -g^{\mu \nu} \partial_{\mu}
\phi \partial_{\nu} \phi = 1 \label{FOURTEEN}, \ee which will be
worked as a constraint on the system. As emphasized in
Sec.~\ref{sec:prelim}, Eq.~(\ref{FOURTEEN}) is a four-dimensional
covariant condition; using the ADM decomposition
(\ref{TWO}) together with the normal derivative
$\nabla_n\phi=(1/N)(\partial_t\phi-N^i\partial_i\phi)$, it is
equivalent to
\be
(\nabla_n\phi)^2 = 1 + h^{ij}\partial_i\phi\,\partial_j\phi,
\label{FIFTEEN}
\ee
which retains the normal derivative of $\phi$ on the same footing
as its spatial gradients [cf. Eq.~(\ref{eq:ADMmimetic})]. The
spatial-only form $-h^{ij}\partial_i\phi\,\partial_j\phi=1$ that
appeared in some intermediate steps of earlier presentations is
algebraically inconsistent on a Riemannian slice, since the
left-hand side is non-positive while the right-hand side is
positive; it is replaced everywhere below by the full ADM form
(\ref{FIFTEEN}).

It is important to point out that the time is taken as a continuous variable and we are working with only spatial variables for a fixed time as has been done in the lattice framework \cite{Wen:2017usd, Fradkin:1978th}. In other words, this is a quasi-static representation of the main situation. Thus we will not consider time evolution of the theory on the lattice as we mentioned in the introduction. Because the construction is canonical, however, the fixed-time data include both the configuration variables and their conjugate momenta; the normal derivatives of the continuum fields are therefore not discarded, but are represented through the canonical momenta, in particular $p_\phi = \sqrt{h}\,\lambda\,\nabla_n\phi$.

For the term $\mathcal{H}^{\circ}_T$ in Eq. (\ref{FOUR}), and the
term $\mathcal{H}_i$ in equation (\ref{THREE}) we will use  \be
\mathcal{H}^{\phi}_T = {1 \over {2 \sqrt{h} \lambda}}p^2_{\phi} + {1
\over 2}\sqrt{h} \lambda (1 + h^{ij}\partial_i \phi \partial_j
\phi), \label{SIXTEEN} \ee and \be \mathcal{H}_i = p_{\phi}
\partial_i \phi - 2 h_{ij} D_k \pi^{jk} \approx 0. \label{SEVENTEEN}
\ee

Now, for the term ${1 \over {2 \sqrt{h} \lambda}}p^2_{\phi}$ in
equation (\ref{SIXTEEN}), it is important to notice that the
discrete form is ${1 \over {2 \sqrt{h} \lambda}} p^2_{\phi_i},$
where we are simply adding an index to define the site location of
the momentum conjugate to the field $\phi_i$. Also, the second term
${1 \over 2}\sqrt{h} \lambda (1 + h^{ij}\partial_i \phi \partial_j
\phi)$ is given by ${1 \over 2}\sqrt{h} \lambda (1 + h_{i, \alpha
\beta} \nabla_{\alpha} \phi_i \nabla_{\beta} \phi_i)$ in the lattice
form.

Equation (\ref{SEVENTEEN}) can be solved to obtain a relation for
the divergence of $\pi^{jk}$: \be D_k \pi^{jk} = {1 \over 2}
p_{\phi} h^{ij} \partial_i \phi, \label{EIGHTEEN} \ee where we are
using the equality symbol for simplicity, given that the constraints
$\mathcal{H}_T \approx 0$, and $\mathcal{H}_i \approx 0$ are
preserved under time evolution, and we will be working on the
reduced phase space.

Equation~(\ref{EIGHTEEN}) is the central continuum statement of the
mimetic sector: the scalar field sources the gravitational momentum
constraint with the current
$J^{j}_{\rm mim}\equiv (1/2)\, p_\phi\, h^{ji}\partial_i\phi$,
which depends explicitly on the canonical momentum $p_\phi$ and
hence, through $p_\phi=\sqrt{h}\,\lambda\,\nabla_n\phi$, on the
normal derivative of $\phi$. Its lattice image, derived next, is
the analogous source term for the fcc rank-two Gauss law
(\ref{TEN}).

If we move this system to the lattice, we can see that equation
(\ref{EIGHTEEN}) can be written as: \be \nabla_{\alpha} \pi_{i,
\beta \alpha} = {1 \over 2} p_{\phi_i} h_{i,\gamma \beta}
\nabla_{\gamma} \phi_i, \label{NINETEEN} \ee where the variables
$\phi_i$, and $p_{\phi_i}$ are defined on the site $i$. This is
precisely a violation of the constraint in equation (\ref{TEN}), and
it appears as predicted by references
\cite{Xu:2010eg,Xu:2006faa,Lozano:2019oxi,Gu:2009jh}.

As a sanity check, taking the long-wavelength limit with lattice
spacing $a$ and the normalized finite difference
\be
\nabla_\alpha f_i \;\longrightarrow\; \partial_\alpha f(x)
+ \mathcal{O}(a),
\label{eq:longwave}
\ee
Eq.~(\ref{NINETEEN}) reduces in a locally Cartesian frame to
$\partial_\alpha \pi_{\beta\alpha} = (1/2)\,p_\phi\,
h_{\gamma\beta}\,\partial_\gamma\phi + \mathcal{O}(a)$, which is
the weak-field form of the continuum momentum-constraint relation
(\ref{EIGHTEEN}). Curvature and connection corrections are
recovered by promoting $\partial_k$ to the covariant derivative
$D_k$ in the continuum expression.

The sourced Gauss law (\ref{NINETEEN}) has an immediate physical
consequence that goes beyond the formal dictionary: it imposes
selection rules on the allowed defect configurations. Writing
$J^{\rm mim}_{i,\beta} \equiv {1 \over 2} p_{\phi_i}
h_{i,\gamma\beta} \nabla_{\gamma}\phi_i$ for the lattice source and
summing Eq.~(\ref{NINETEEN}) over all sites of a periodic lattice
$\Lambda$, each discrete derivative telescopes,
\be
Q_\beta \;\equiv\; \sum_{i\in\Lambda} J^{\rm mim}_{i,\beta}
\;=\; \sum_{i\in\Lambda} \nabla_\alpha \pi_{i,\beta\alpha}
\;=\; 0,
\label{eq:chargerule}
\ee
so the total mimetic charge vanishes identically. On a lattice with
boundary $\partial\Lambda$ the same telescoping gives
$Q_\beta = \sum_{\partial\Lambda} \pi_{\beta\alpha}\, n^\alpha$,
with $n^\alpha$ the outward lattice normal. Furthermore, since
$\pi_{i,\beta\alpha}$ is symmetric, the antisymmetrized first
moment of the source obeys a second identity: for field
configurations for which the weighted boundary term
$\sum_{\partial\Lambda} (x_\alpha \pi_{\beta\gamma} - x_\beta
\pi_{\alpha\gamma})\, n^\gamma$ vanishes (for example, when
$\pi_{\alpha\beta}$ has compact support or sufficiently rapid
asymptotic decay, or when this term vanishes by symmetry, as it
does for the isotropic solution of
Sec.~\ref{subsec:pointdefect}), summation by parts with
$\nabla_\gamma x_\alpha = \delta_{\gamma\alpha}$ yields
\be
\sum_{i}
\left( x_\alpha J^{\rm mim}_{i,\beta}
- x_\beta J^{\rm mim}_{i,\alpha} \right)
= - \sum_{i}
\left( \pi_{i,\beta\alpha} - \pi_{i,\alpha\beta} \right)
= 0.
\label{eq:momentrule}
\ee
Physically, Eqs.~(\ref{eq:chargerule})--(\ref{eq:momentrule}) state
that on a closed lattice a single unbalanced mimetic defect cannot
exist in isolation: mimetic sources must appear in globally neutral
configurations, with vanishing net charge and vanishing net
antisymmetric moment, or else be compensated by flux through the
boundary. These are the rank-two analogs of the familiar charge
selection rules of lattice electrodynamics, and they parallel the
constrained charge sectors of higher-rank gauge theories
\cite{Pretko:2017xar,You:2024prv}. In this precise sense the
construction does more than translate the continuum source into
lattice language: it classifies which lattice configurations can
carry a mimetic dark-sector inhomogeneity at all.

\begin{figure}
\begin{center}
\includegraphics[scale=0.6]{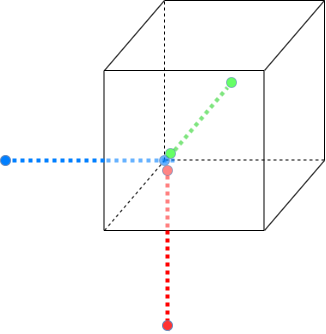}
\caption[Short caption for figure 4]{\label{Figure4} {\scriptsize Representation of $\nabla_{\gamma} \phi_i$ in the links $(i - \gamma, \gamma)$, with $\gamma = x$ in blue, $\gamma = y$ in red, and $\gamma = z$ in green. Physically, these links carry the spatial-gradient factor of the mimetic source $J^{\rm mim}_{i,\beta}$ in Eq.~(\ref{NINETEEN}): the scalar lives on sites and its lattice gradient on links, so the mimetic momentum density enters the Gauss law as a link-supported charge.}}
\end{center}
\end{figure}

\begin{figure}
\begin{center}
\includegraphics[scale=0.6]{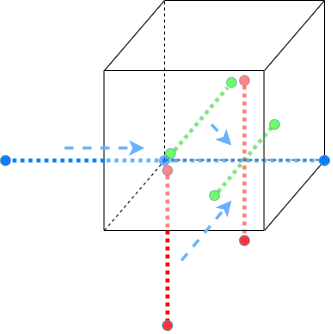}
\caption[Short caption for figure 5]{\label{Figure5} {\scriptsize Displacement of the term $\nabla_{\gamma} \phi_i$ by the term ${1 \over 2} p_{\phi_i} h_{i,\gamma \beta}$ in the product of equation (\ref{NINETEEN}). Physically, the multiplication by ${1 \over 2} p_{\phi_i} h_{i,\gamma\beta}$ transports the scalar-gradient links onto the plaquette positions where the rank-two source must be supported for the sourced Gauss law to close site by site; this is the lattice image of contracting the continuum source with the spatial metric.}}
\end{center}
\end{figure}

We can observe that equation (\ref{NINETEEN}) is an equation of
vectors in the direction $\beta$. For the right-hand-side, the terms
$\nabla_{\gamma} \phi_i$ can be seen as the links $(i, \gamma)$, but
we will use backward differentiation instead of the forward
differentiation for matters of representation, so the links are $(i
- \gamma, \gamma)$. These terms are gathered in Figure \ref{Figure4}
for $\beta = x$. In equation (\ref{NINETEEN}), $\nabla_{\gamma}
\phi_i$ is multiplied by ${1 \over 2} p_{\phi_i} h_{i,\gamma
\beta}$, which can be regarded as moving all the link ${1 \over 2}$
units in the $\gamma$ direction and ${1 \over 2}$ units in the
$\beta$ direction. This displacement can be observed in Figure
\ref{Figure5} for $\beta = x$.

\begin{figure}
\begin{center}
\includegraphics[scale=0.6]{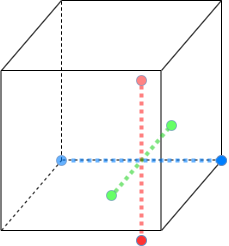}
\caption[Short caption for figure 6]{\label{Figure6} {\scriptsize Representation of $\nabla_{\alpha} \pi_{i, \beta \alpha}$ with backward differentiation in $y$, and $z$, but forward differentiation in $x$, interpreted as a remnant of the covariant derivative. Physically, this pattern is the local support of the discrete divergence entering the Gauss law: a nonzero mimetic source appears as a local failure of the vacuum rank-two Gauss law (\ref{TEN}) on the indicated cell, with the same link geometry as the constraint term of Fig.~\ref{Figure3}, which is what identifies the source as a defect of that constraint.}}
\end{center}
\end{figure}

For the left-hand-side of equation (\ref{NINETEEN}), we can see that
the result has to be the same as the right-hand-side (see Figure
\ref{Figure6} for the case $\beta = x$), so fitting it we observe
that the terms $\nabla_y \pi_{i, xy}$, and $\nabla_z \pi_{i, xz}$
keep using the backward differentiation, but the term $\nabla_x
\pi_{i, xx}$ is using the forward differentiation. This could be
interpreted as a remnant of the covariant derivative, when
translated into the lattice. The congruence of the term in equation
(\ref{NINE}), and the $\beta = x$ term of equation (\ref{NINETEEN})
is visible in Figures \ref{Figure3} and \ref{Figure6}. For $\beta =
y$, $z$ we have a similar situation.

The lattice equation (\ref{NINETEEN}) is the discrete image of the
continuum momentum-constraint relation (\ref{EIGHTEEN}), which
itself follows from the spatial diffeomorphism constraint
(\ref{SEVENTEEN}). The Lagrange multiplier $\lambda$ in the
Hamiltonian (\ref{SIXTEEN}) is fixed on the constraint surface by
its own variation. Concretely, preservation of the primary
constraint $p_\lambda\approx 0$ yields the secondary condition
\be
-\frac{p_\phi^2}{2\sqrt{h}\,\lambda^2}
+ \frac{\sqrt{h}}{2}
\left(1 + h^{ij}\partial_i\phi\,\partial_j\phi\right)
\approx 0,
\label{eq:lambdasecondary}
\ee
equivalently
$-p_\phi^2/(h\lambda^2) + 1 + h^{ij}\partial_i\phi\,\partial_j\phi
\approx 0$. Using the canonical relation
$p_\phi=\sqrt{h}\,\lambda\,\nabla_n\phi$, this is precisely the
ADM mimetic constraint (\ref{FIFTEEN}), so $\lambda$ is determined
on the constraint surface rather than treated as an external large
parameter (cf.\ \cite{Chaichian:2014qba,Ganz:2018mqi}). The
geometric content of the term
$(1+h_{i,\alpha\beta}\nabla_\alpha\phi_i\nabla_\beta\phi_i)$ in
Eq.~(\ref{SIXTEEN}) is then just the spatial part of the same ADM
constraint, and the relationship between $\lambda$ and a second
multiplier $\omega$ appearing once a dark-energy companion field
is added will be analyzed in Sec.~\ref{subsec:inhodarkener}.

\subsection{A localized defect and its long-range field}
\label{subsec:pointdefect}

The selection rules (\ref{eq:chargerule})--(\ref{eq:momentrule})
constrain which defect configurations are admissible. Now we will
show, in a worked example, that an idealized localized source
(understood as carrying compensating flux at infinity, or as one
member of a globally neutral configuration, consistently with
those rules) is not an inert label: it forces a long-range
field response of the rank-two sector, and the trace-free
representative of this response is exactly the Bowen--York
momentum solution of canonical gravity. We work at long
wavelength, where by Eq.~(\ref{eq:longwave}) the lattice problem
reduces to a continuum one, and we comment on lattice corrections
at the end.

Consider a static point source $J_\beta\,\delta^3(x)$ for the
constraint, with $J_\beta$ the mimetic charge of the defect [the
lattice source $J^{\rm mim}_{i,\beta}$ concentrated on one site].
The constraint alone does not determine $\pi_{\beta\alpha}$. For
this worked example we specialize to the rotationally invariant
quadratic continuum limit of the effective Hamiltonian
(\ref{ELEVEN}), identifying the diagonal and off-diagonal
stiffnesses after normalization, $E = \frac{1}{2K}\int d^3x\,
\pi_{\beta\alpha}\pi_{\beta\alpha}$, and we minimize this energy
at fixed $h_{\alpha\beta}$ subject to
$\partial_\alpha \pi_{\beta\alpha} = J_\beta\,\delta^3(x)$. A
natural representative of the sourced sector is the configuration
minimizing this isotropic energy.
Introducing a vector Lagrange-multiplier field for the constraint,
the minimizer is a symmetrized gradient, and in Fourier space the
transverse and longitudinal parts of the multiplier are fixed
separately, giving
\be
\pi_{\beta\alpha}(k)
= -\,i\,\frac{k_\alpha J_\beta + k_\beta J_\alpha}{k^2}
+ i\,\frac{k_\alpha k_\beta\, (k\cdot J)}{k^4},
\label{eq:pointk}
\ee
which satisfies $i k_\alpha \pi_{\beta\alpha}(k) = J_\beta$
exactly. Transforming to position space, with $n_\alpha =
x_\alpha/r$,
\be
\pi_{\beta\alpha}(x)
= \frac{\frac{1}{2}\left(n_\alpha J_\beta + n_\beta J_\alpha\right)
+ \frac{3}{2}\, n_\alpha n_\beta\, (n\cdot J)
- \frac{1}{2}\,\delta_{\alpha\beta}\, (n\cdot J)}{4\pi r^2}.
\label{eq:pointr}
\ee
The field of the defect therefore falls off as $1/r^2$, the
rank-two analog of the Coulomb law, and its flux through any
sphere enclosing the defect reproduces the total charge,
$\oint_{S^2} \pi_{\beta\alpha}\, n_\alpha\, r^2 d\Omega = J_\beta$,
in agreement with the boundary form of the selection rule
(\ref{eq:chargerule}).

Equation (\ref{eq:pointr}) is the minimum-energy representative
when no condition is imposed on the trace; indeed its trace is
$\pi^{\alpha}{}_{\alpha} = (n\cdot J)/(4\pi r^2)$. In canonical
general relativity the analogous object is usually taken
trace-free, corresponding to the maximal-slicing condition. It is
important to point out that the same construction gives that
representative as well. Adding to Eq.~(\ref{eq:pointk}) the
divergence-free correction
$$
H_{\alpha\beta}(k)
= \frac{i}{2}
\left( \delta_{\alpha\beta} - \frac{k_\alpha k_\beta}{k^2} \right)
\frac{k\cdot J}{k^2},
$$
which carries no source and cancels the trace of
Eq.~(\ref{eq:pointk}), one obtains the trace-free representative
of the same sourced sector,
\be
\pi^{\rm TF}_{\alpha\beta}(x)
= \frac{3}{16\pi r^{2}}
\left[ n_\alpha J_\beta + n_\beta J_\alpha
- \left( \delta_{\alpha\beta} - n_\alpha n_\beta \right)
(n\cdot J) \right],
\label{eq:pointrTF}
\ee
with the same flux $J_\beta$. Since the energy is convex and both
constraints (the sourced divergence and the vanishing trace) are
linear, this is the unique minimizer within the trace-free sourced
sector, once the problem is regulated in the infrared as it is on
any finite lattice. Now, identifying $J_\alpha = 8\pi P_\alpha$,
Eq.~(\ref{eq:pointrTF}) becomes exactly
\be
\widetilde{A}^{ij}_{\rm BY}
= \frac{3}{2 r^{2}}
\left[ P^{i} n^{j} + P^{j} n^{i}
- \left( \delta^{ij} - n^{i} n^{j} \right) (P\cdot n) \right],
\label{eq:BY}
\ee
the Bowen--York solution of the momentum constraint of general
relativity \cite{Bowen:1980yu}: the standard conformal trace-free
solution carrying ADM linear momentum, normalized so that $P^i =
(1/8\pi)\oint \widetilde{A}^{ij}_{\rm BY} n_j\, dA$, and widely
used for boosted black-hole puncture initial data. Bowen--York
data solve the vacuum constraint on punctured $\mathbb{R}^3$;
extended distributionally over all of $\mathbb{R}^3$, the puncture
acts as a delta-function momentum source \cite{Tonita:2011}, which
is exactly the problem solved above. Equation (\ref{eq:BY}) is
written as the conformal trace-free extrinsic curvature
$\widetilde{A}^{ij}$, the standard object of the
numerical-relativity literature. In the canonical normalization of
Sec.~\ref{sec:prelim}, where $\pi^{ij} = {1 \over 2}\sqrt{h}
(K^{ij} - h^{ij} K)$, one has $\pi^{ij} = {1 \over 2}
\widetilde{A}^{ij}$ on a flat maximal slice, so the same identity
expressed in terms of the canonical momentum corresponds to
$J_\alpha = 4\pi P_\alpha$. In other words: the mimetic defect and a point particle
carrying ADM momentum source the same constraint sector, and the
trace-free representative of the defect field is the Bowen--York
tensor itself. In this precise constraint-level sense, the defect
sources nontrivial gravitational canonical momentum.

Two remarks complete the example. First, combined with the
selection rules above, the long-range response indicates that the
defect sector of the lattice model carries restricted dynamics of
the kind familiar from higher-rank gauge theories, where
conservation of charges and of certain moments constrains the
mobility of excitations \cite{Pretko:2017fbf,Pretko:2017xar,
You:2024prv}; we do not claim, however, that the mimetic defects
are fractons in the technical sense without an analysis of the
allowed hopping operators. Second, on the lattice
Eq.~(\ref{eq:pointr}) receives $\mathcal{O}(a)$ corrections
through Eq.~(\ref{eq:longwave}); moreover, the energy of the
point-source configuration diverges linearly at the core, $E \sim
1/a$, and on the lattice this divergence is regulated
automatically by the lattice spacing, which acts as the core
regulator. Also, for generic stiffnesses
$K_1 \neq K_2$ in Eq.~(\ref{ELEVEN}) the source still produces a
long-range elliptic response, but its angular profile carries
cubic anisotropy, and the exact Bowen--York structure is
recovered only in the isotropic limit considered here. Finally,
the minimal-energy profile is one representative of the sourced
sector: any two solutions of the sourced Gauss law differ by a
solution of the vacuum constraint (\ref{TEN}).

\subsection{Inhomogeneous Dark Energy Model}
\label{subsec:inhodarkener}

Now, we perform a different modification to equation (\ref{ONE}), by
adding only one extra scalar field $\psi$ to the mimetic dark matter
model of Section \ref{subsec:mimedarkmatt} in
\cite{Chamseddine:2016uyr, Kluson:2017iem}. In this way, we end up
with the following relations: \be -g^{\mu \nu}
\partial_{\mu} \phi \partial_{\nu} \phi = 1, \qquad \quad g^{\mu
\nu} \partial_{\mu} \psi \partial_{\nu} \phi = 0. \label{TWENTY} \ee

Here, it is important to point out that in the case of disformal
transformations \cite{Firouzjahi:2018xob}, the second constraint
equation in (\ref{TWENTY}) does not has to hold. But as defined in
\cite{Bekenstein:1992pj} and observed in \cite{Chamseddine:2013kea},
the transformations we are working with are conformal and they have
to be that way to fit with our purposes, since the added scalar
field $\phi$ is related to the conformal factor $\Phi$ without
adding a new dynamical variable.

These last two relations modify the term $\mathcal{H}_T^{\circ}$ of
equation (\ref{FOUR}) to get: $$ \mathcal{H}^{\phi, \psi}_T = {2
\over \sqrt{h} \omega} p_{\phi} p_{\psi} - {2 \lambda \over \sqrt{h}
\omega^2} p^2_{\psi} + {1 \over 2} \lambda \sqrt{h}(1 + h^{ij}
\partial_i \phi \partial_j \phi) $$ \be + {1 \over 2} \sqrt{h}
\omega h^{ij} \partial_i \psi \partial_j \phi + {1 \over 2} \sqrt{h}
V(\psi), \label{TWENTYONE} \ee and \be \mathcal{H}_i = p_{\phi}
\partial_i \phi + p_{\psi} \partial_i \psi - 2h_{ik} D_j \pi^{kj}
\approx 0, \label{TWENTYTWO} \ee where $V(\psi)$ is a potential term
that can be added to the Hamiltonian, $\lambda$ is a Lagrange
multiplier for the first equation (\ref{TWENTY}), and $\omega$ is a
Lagrange multiplier for the second equation (\ref{TWENTY}).

The lattice form of equation (\ref{TWENTYONE}) makes $p_{\phi}
\rightarrow p_{\phi_i}$, and $p_{\psi} \rightarrow p_{\psi_i}$,
keeps the restriction for $\phi_i$ in the term ${1 \over 2}\sqrt{h}
\lambda (1 + h_{i, \alpha \beta} \nabla_{\alpha} \phi_i
\nabla_{\beta} \phi_i)$, and gives rise to a new restriction on
$\psi_i$, related to the one of $\phi_i$, in the term ${1 \over 2}
\sqrt{h} \omega h_{i, \alpha \beta} \nabla_{\alpha} \psi_i
\nabla_{\beta} \phi_i$. As clarified in the ADM analysis below
[cf.~Eq.~(\ref{TWENTYSIXandahalf})], this last term restricts the
spatial gradients of $\phi_i$ and $\psi_i$ from being parallel only
in the spatial-gradient-orthogonal subsector; in the generic
inhomogeneous case the spatial cross term is balanced by the product
of the two normal derivatives. The potential $V(\psi)$ will be
considered below.

Following the process of Section \ref{subsec:mimedarkmatt}, we can
see that we have \be D_k \pi^{jk} = {1 \over 2} p_{\phi} h^{ij}
\partial_i \phi + {1 \over 2} p_{\psi} h^{ij} \partial_i \psi,
\label{TWENTYTHREE} \ee and moving this term to the lattice, we
obtain \be \nabla_{\alpha} \pi_{i, \beta \alpha} = {1 \over 2}
p_{\phi_i} h_{i,\gamma \beta} \nabla_{\gamma} \phi_i + {1 \over 2}
p_{\psi_i} h_{i,\gamma \beta} \nabla_{\gamma} \psi_i.
\label{TWENTYFOUR} \ee

It is straightforward to see the similarity between equations
(\ref{NINETEEN}) and (\ref{TWENTYFOUR}), but the preservation of the
momenta conjugate to $\lambda$ and $\omega$ in equation
(\ref{TWENTYONE}) are very different. As seen in reference
\cite{Kluson:2017iem}, the preservation of the momentum conjugate to
$\lambda$ gives rise to \be N\bigg({2 \over \sqrt{h} \omega^2}
p_{\psi}^2 - {1 \over 2} \sqrt{h} (1 + h^{ij} \partial_i \phi
\partial_j \phi)\bigg) \approx 0, \label{TWENTYFIVE} \ee and the
preservation of the momentum conjugate to $\omega$ yields \be
N\bigg({2 \over \sqrt{h} \omega^2} p_{\psi} p_{\phi} - {4 \lambda
\over \sqrt{h} \omega^3} p_{\psi}^2 - {1 \over 2} \sqrt{h} h^{ij}
\partial_i \psi \partial_j \phi\bigg) \approx 0. \label{TWENTYSIX} \ee

The second constraint in (\ref{TWENTY}) is a four-dimensional
covariant condition. In ADM form, with the normal derivatives
$\nabla_n\phi$ and $\nabla_n\psi$ defined as in
Sec.~\ref{sec:prelim}, it reads
\be
-(\nabla_n\psi)(\nabla_n\phi)
+ h^{ij}\partial_i\psi\,\partial_j\phi = 0,
\label{TWENTYSIXandahalf}
\ee
so that the spatial-gradient cross term
$h^{ij}\partial_i\psi\,\partial_j\phi$ does not vanish in general
but is fixed by the product of the two normal derivatives.
For the spatial-gradient-orthogonal subsector
$h^{ij}\partial_i\psi\,\partial_j\phi = 0$ (equivalently
$(\nabla_n\psi)(\nabla_n\phi)=0$), the relation extracted from
Eq.~(\ref{TWENTYSIX}) reduces to
\be
{p_{\phi} \over p_{\psi}} = {2\lambda \over \omega}.
\label{TWENTYSEVEN}
\ee
Outside this subsector, the same manipulation of
Eq.~(\ref{TWENTYSIX}) yields the corrected relation
\be
{p_{\phi} \over p_{\psi}}
=
{2\lambda \over \omega}
+ {h\,\omega^2 \over 4 p_{\psi}^2}\,
h^{ij}\partial_i\psi\,\partial_j\phi,
\label{TWENTYSEVENb}
\ee
which is therefore the form to be used in the generic
inhomogeneous case. Both $\lambda$ and $\omega$ are determined on
the constraint surface by their respective variations, together
with the two-field canonical relations that follow from
Eq.~(\ref{TWENTYONE}),
\be
p_\psi = \frac{\sqrt{h}\,\omega}{2}\,\nabla_n\phi,
\qquad
p_\phi = \frac{\sqrt{h}\,\omega}{2}\,\nabla_n\psi
+ \lambda\,\sqrt{h}\,\nabla_n\phi.
\label{eq:twofieldp}
\ee
The simple one-field relation
$p_\phi=\sqrt{h}\,\lambda\,\nabla_n\phi$ of
Sec.~\ref{subsec:mimedarkmatt} should therefore not be imported
unchanged into the two-field sector. The physically meaningful
quantity is the ratio $p_{\phi}/p_{\psi}$ given by
Eqs.~(\ref{TWENTYSEVEN})--(\ref{TWENTYSEVENb}); it controls the
relative weight of the two scalar contributions in the
momentum-constraint source. The parametric classification
$$\lambda \ll \omega, \qquad \lambda \sim \omega, \qquad
\lambda \gg \omega$$
should be read as a labeling of distinct sectors of the reduced
phase space rather than as a limit on either multiplier
separately.

In the spatial-gradient-orthogonal subsector, or more generally
when the correction term in Eq.~(\ref{TWENTYSEVENb}) is subleading,
the condition $\lambda \ll \omega$ implies $p_{\phi_i} \ll
p_{\psi_i}$, making the field $\psi_i$ more dynamical than the
field $\phi_i$. Note that the first restriction in (\ref{TWENTY})
constrains the field $\phi_i$ through the metric $h_{i,\alpha\beta}$
from one site to another in the vicinity. In the controlled
spatial-gradient-orthogonal subsector, the second restriction does
not force one of the two fields to be constant; rather, it
constrains their spatial gradients to be orthogonal, or, more
generally, makes the cross-gradient correction in
Eq.~(\ref{TWENTYSEVENb}) subleading. Thus both $\phi_i$ and
$\psi_i$ may be present on the same site, while their relative
contribution to the source (\ref{TWENTYFOUR}) is controlled by
$p_{\phi_i}/p_{\psi_i}$. For $\lambda \ll \omega$, the $\psi_i$
contribution dominates the lattice source, and the divergence law
in Eq.~(\ref{TWENTYFOUR}) for the metric conjugate momentum can be
understood as relying mostly on the dark-energy-like component.

Within the same regime of validity, as one turns $\lambda \sim
\omega$, this makes $p_{\phi_i} \sim p_{\psi_i}$, equilibrating
them in equation (\ref{TWENTYFOUR}), and also balances them.
Finally, again within this controlled subsector, when $\lambda \gg
\omega$, we also have $p_{\phi_i} \gg p_{\psi_i}$, giving rise to
a more dynamical dark matter than dark energy. This regime can be
regarded as a pure dark-matter sector, like the one from the last
section, in the case in which $V(\psi_i) = 0$. If $V(\psi_i) \neq 0$, then the last two
terms of equation (\ref{TWENTYONE}) can be combined, and we can see
the dependence of this potential directly on ${1 \over 2} \sqrt{h}
\omega h_{i, \alpha \beta} \nabla_{\alpha} \psi_i \nabla_{\beta}
\phi_i$. Notice that this is the only way in which the potential $V$
can be added in our Hamiltonian, or combined with all the terms
depending on $\psi_i$, or $p_{\psi_i}$.

Following these results, it can be observed that $\lambda$ and
$\omega$ play the role of regime parameters separating the
dark-matter-dominated and dark-energy-dominated sectors of the
two-scalar source. A genuine phase-transition interpretation
would require an effective potential analysis along the lines of
the recent multi-field mimetic studies
\cite{Shen:2019nyp,Mansoori:2021bjj,Zheng:2022rad}, in which the
two-field mimetic system is shown to carry two propagating scalar
degrees of freedom whose stability depends on the signature of the
field-space metric. Within that framework, the recent two-field
mimetic literature interprets the mimetic energy density as
predominantly dark-matter-like at the background level, with
genuine dark-energy behavior requiring additional ingredients
such as scalar potentials, curved field space, or higher-derivative
mimetic extensions. We therefore use ``inhomogeneous dark energy''
here to refer to the inhomogeneous structure of the lattice
source, rather than as a claim that the minimal two-scalar
mimetic completion realizes a viable cosmological dark-energy
background; that stronger claim would require the extra structure
discussed in those references.

Closely related considerations have been discussed in the context
of dark-sector unification through the dark-energy density alone
\cite{Wang:2004ru} and through a single scalar describing both
dark matter and dark energy \cite{Brandenberger:2018xnf,
Brandenberger:2019jfh}. If the density of dark energy genuinely
varies with time, one could use $\omega$ as a parameter linked to
the time, which would give a connection between our quasi-static
model and a dynamical one.

\subsection{Vector Field Mimetic Dark Matter Model}
\label{subsec:vectormimedarkmatt}

Once again, following the line of \cite{Chaichian:2014qba} and
modifying equation (\ref{ONE}) by putting the physical metric
$g^{\textrm{phys}}_{\mu \nu}$ in terms of a vector field
$u_{\alpha},$ and the fundamental metric $g_{\mu \nu},$ with the
same gauge fixing, $\Phi - 1 = 0$, we obtain: \be -g^{\alpha \beta}
u_{\alpha} u_{\beta} = 1. \label{TWENTYEIGHT} \ee

There are some modifications for the term $\mathcal{H}^{\circ}_T$ of
the Hamiltonian density in equation (\ref{FOUR}), for which we have
$$ \mathcal{H}^{u}_T =  -{1 \over 2} \sqrt{h} \lambda (1 + h^{ij}
u_i u_j - u^2_{\textbf{n}}) + {1 \over 2 \mu^2 \sqrt{h}} h_{ij}
p^{i} p^j $$ \be + {\mu^2 \over 4} \sqrt{h} h^{ik} h^{jl} (D_i u_j -
D_j u_i)(D_k u_l - D_l u_k) - u_{\textbf{n}} D_i p^{i},
\label{TWENTYNINE} \ee and \be \mathcal{H}_i = \partial_i u_j p^j -
\partial_j(u_i p^j) - 2h_{ik} D_j \pi^{kj} \approx 0, \label{THIRTY}
\ee where $p_i$ is the momentum conjugate to the vector field $u_i,$
and $u_{\textbf{n}}$ is the component of the vector field normal to
the constant-time surfaces. The component $u_{\textbf{n}}$ is
fixed algebraically by the unit-timelike constraint
$-g^{\alpha\beta}u_\alpha u_\beta = 1$ of
Eq.~(\ref{TWENTYEIGHT}), in direct analogy with the way the scalar
mimetic constraint (\ref{FIFTEEN}) determines $\nabla_n\phi$ in
Sec.~\ref{subsec:mimedarkmatt}; the dynamical degrees of freedom
in the lattice formulation below are therefore the spatial
components $u_{i,\alpha}$, while $u_{\textbf{n}}$ enters through
the constraint and through the coupling term $u_{\textbf{n}} D_i
p^i$. In what follows we work on the constraint surface and write
the lattice Hamiltonian in terms of $u_{i,\alpha}$ and
$p_{i,\alpha}$, with $u_{\textbf{n}}$ playing the role of a
site-local scalar parameter on the lattice. The construction is a
discrete analogue of the canonical vector-mimetic formulation of
\cite{Chaichian:2014qba} and of the gauge-field mimetic
cosmologies analyzed in \cite{Gorji:2018okn,Gorji:2019ttx}.

In order to write equation (\ref{TWENTYNINE}) in a lattice form we
observe that the first term, the lattice image of the
unit-timelike constraint $1 + h_{i,\alpha\beta}u_{i,\alpha}u_{i,\beta}
- u_{\textbf{n} i}^{\,2} = 0$, is a constraint on the amount of
dimers defined surrounding the site $i$ and contains the algebraic
dependence on $u_{\textbf{n} i}$ through the squared term. In
addition, $u_{\textbf{n}}$ enters the Hamiltonian through the
coupling term $u_{\textbf{n}} D_i p^i$. Variation of
Eq.~(\ref{TWENTYNINE}) with respect to $u_{\textbf{n}}$, taken
together with the unit-timelike constraint (\ref{TWENTYEIGHT}),
relates the longitudinal part of the vector momentum to
$\lambda$ and $u_{\textbf{n}}$; explicitly, $D_i p^i$ is sourced
by $\sqrt{h}\,\lambda\,u_{\textbf{n}}$ up to convention. Thus
$u_{\textbf{n}}$ controls the divergence of $p^i$ rather than
forcing it to vanish, and the lattice divergence
$\nabla_\alpha p_{i,\alpha}$ should be interpreted as the
vector-sector source entering the Gauss-law defect derived in
Eq.~(\ref{THIRTYTWO}) below, in direct analogy with the role
played by $\nabla_\gamma\phi_i$ in the scalar case of
Sec.~\ref{subsec:mimedarkmatt}.

The kinetic term ${1 \over 2 \mu^2 \sqrt{h}} h_{i, \alpha \beta}
p_{i, \alpha} p_{i, \beta}$ is suppressed by $\mu^2$. In the
strong-coupling or low-energy sector in which the magnetic
contribution proportional to $\mu^2 (D_i u_j - D_j u_i)^2$ is
energetically suppressed, one may project onto approximately
curl-free vector configurations, $D_i u_j - D_j u_i \simeq 0$,
which on the lattice translates to $\nabla_\gamma u_{i,\delta}
\simeq \nabla_\delta u_{i,\gamma}$. We emphasize that this is an
energetic restriction rather than the result of sending
$\mu\to\infty$ as a Lagrange-multiplier-type limit, and that for
finite $\mu$ the momentum term remains in the lattice
Hamiltonian. When the divergence $\nabla_\alpha p_{i, \alpha}$ is
non-zero in this regime, it sources the vector contribution to
the Gauss-law defect and may be read as a torsion-like ingredient
for $u_{i, \alpha}$ in the corresponding continuum-limit
interpretation.

For the constraint in equation (\ref{THIRTY}), under the
approximate curl-free condition $\nabla_\gamma u_{i,\delta}\simeq
\nabla_\delta u_{i,\gamma}$ identified above, the term
$\partial_i u_j p^j - \partial_j (u_i p^j) = (\partial_i u_j -
\partial_j u_i) p^j - u_i \partial_j p^j$ reduces to $- u_i
\partial_j p^j$, or in lattice form $-u_{i, \gamma}
\nabla_{\delta} p_{i, \delta}$. The form of the resulting
defect is therefore the same whether $u_{\textbf{n} i} = 0$ or
not. With
this, we can follow the steps of Section \ref{subsec:mimedarkmatt},
and obtain a modification of equation (\ref{THIRTY}) to be: \be D_j
\pi^{kj} = - {1 \over 2} h^{ik} u_i \partial_j p^j
\label{THIRTYONE}, \ee which in lattice form can be written as: \be
\nabla_{\alpha} \pi_{i, \beta \alpha} = - {1 \over 2} h_{i, \gamma
\beta} u_{i, \gamma} \nabla_{\delta} p_{i, \delta}
\label{THIRTYTWO}. \ee

It can be observed from this equation that the divergence of
$\pi_{i, \beta \alpha}$ is directly related to divergence of $p_{i,
\delta}$, which is given by a scalar quantity defined at the site
$i$. This is achieved by using the minus sign with the forward
differentiation. Then, this scalar quantity is extended along the
direction $\beta$ by the projection of $h_{i, \gamma \beta} u_{i,
\gamma}.$ The final result is a vector (dimer) in the link $(i,
\beta)$, like in the last two sections.

A few remarks on the relativistic regime of the construction are
in order. The fcc rank-two lattice graviton model on which our
construction is built is a Lifshitz-type emergent-gravity model
in the sense of \cite{Xu:2010eg}, and we do not derive the
real-time propagation of tensor modes from it. The
post-GW170817 bound $|c_T-1|\lesssim 10^{-15}$ on the
gravitational-wave speed therefore acts in our setting as a
condition on the continuum mimetic-vector completion to which
the lattice map is applied, rather than as a result of the
present construction: any continuum completion of the vector
sector that is to provide a phenomenologically viable
gravitational theory must independently satisfy this bound, and
we restrict attention to such completions
\cite{Casalino:2018tcd,GBM:2017lvd,Sakstein:2017xjx}. The vector
sector is otherwise constrained by the disformal multi-field
analyses of \cite{Firouzjahi:2018xob}, and the residual stability
issues are addressed jointly with the scalar sector in
Sec.~\ref{sec:stability-scope}.

\subsection{Mimetic Tensor-Vector-Scalar Models}
\label{subsec:mimeteves}

After GW170817, relativistic modified-gravity models with
non-luminal tensor propagation, including many realizations of
the Tensor-Vector-Scalar (TeVeS) framework, became severely
constrained by the bound $|c_T-1|\lesssim 10^{-15}$
\cite{Casalino:2018tcd,Sakstein:2017xjx}. Mimetic
tensor-vector-scalar constructions of the type considered here
can be arranged to satisfy this bound and continue to be
developed as phenomenological models of the dark sector
\cite{Chaichian:2014qba,Chaichian:2014dfa,Benisty:2021cmq}. A
prominent relativistic-MOND program that has emerged in the
same period is the Aether-Scalar-Tensor (AeST) proposal of
Skordis and Zlosnik \cite{Skordis:2020eui,Skordis:2021ahj}, in
which a unit-timelike vector and a scalar field are coupled to
gravity in a way designed to recover the MOND limit while
remaining viable after GW170817. Dom\`enech and Ganz
\cite{Domenech:2025bqe} have furthermore shown that
relativistic-MOND theories featuring a unit-timelike vector,
such as TeVeS and AeST, can be embedded in a conformal-disformal
framework admitting a mimetic representation, so that the lattice
realization studied here is naturally a discrete image of a
still active continuum thread.

As presented in \cite{Chaichian:2014qba, Chaichian:2014dfa}, we can
modify equation (\ref{ONE}) by keeping the same gauge fixing  ($\Phi
- 1 = 0$) and writing the physical metric $g^{\textrm{phys}}_{\mu
\nu}$ in terms of a scalar field $\phi$, and a vector field
$u_{\mu}$: \be -f(\phi) (g^{\alpha \beta} u_{\alpha} u_{\beta}) = 1,
\label{THIRTYTHREE} \ee with $f(\phi)$ a nonnegative function of
$\phi$.

The corresponding modifications to the term $\mathcal{H}^{\circ}_T$
of equation (\ref{FOUR}) are given by: $$ \mathcal{H}^{u, \phi}_T =
-{1 \over 2} \sqrt{h} \lambda (1 + f(\phi) h^{ij} u_i u_j - f(\phi)
u^2_{\textbf{n}}) + {1 \over 2 \mu^2 \sqrt{h}} h_{ij} p^{i} p^j $$
$$ + {\mu^2 \over 4} \sqrt{h} h^{ik} h^{jl} (D_i u_j - D_j u_i)(D_k
u_l - D_l u_k) - u_{\textbf{n}} D_i p^{i} $$ \be + {p^2_{\phi} \over
2 \sqrt{h}} + {1 \over 2} \sqrt{h} h^{ij} \partial_i \phi \partial_j
\phi + V(\phi),  \label{THIRTYFOUR} \ee and \be \mathcal{H}_i =
p_{\phi} \partial_i \phi + \partial_i u_j p^j - \partial_j(u_i p^j)
- 2h_{ik} D_j \pi^{kj} \approx 0, \label{THIRTYFIVE} \ee where
$V(\phi)$ is a potential term that can be added to the Hamiltonian.

The lattice form of equation (\ref{THIRTYFOUR}), is the same as in
the last section, but the function $f(\phi_i)$ makes a variation on
how $u_{i, \alpha}$ changes depending on the site, the scalar
$\phi_i$, and the function $f(\phi_i)$ defined on it. As observed
before, the divergence of $p_{i, \alpha}$ gives rise to a divergence
on $\pi_{i, \alpha \beta}$, and this will now have a balance with
the gradient of $\phi_i$ (see equation (\ref{THIRTYSIX}) below). The
last three terms of equation (\ref{THIRTYFOUR}) constitute the
scalar contribution already present in the mimetic-scalar sector;
they do not alter the vector-dimer structure except through the
scalar source $p_{\phi_i}h_{i,\gamma\beta}\nabla_\gamma\phi_i$
appearing in Eq.~(\ref{THIRTYSIX}).

Following the steps of Sections \ref{subsec:mimedarkmatt} and
\ref{subsec:vectormimedarkmatt} we can obtain a discrete model for
the constraint in equation (\ref{THIRTYFIVE}) as \be \nabla_{\alpha}
\pi_{i, \beta \alpha} = {1 \over 2} p_{\phi_i} h_{i, \gamma \beta}
\nabla_{\gamma} \phi_{i}  - {1 \over 2} h_{i, \gamma \beta} u_{i,
\gamma} \nabla_{\delta} p_{i, \delta} \label{THIRTYSIX}, \ee which
gives rise to a dimer in case any of the elements in the right side
is different from zero.

\subsection{Stability and scope of the construction}
\label{sec:stability-scope}

We first state the dynamical status of the correspondence
established in the preceding subsections. The correspondence is a
constraint-level correspondence: it maps the mimetic contribution
to the ADM momentum constraint into a source of the rank-two
lattice Gauss law, exactly at the level of the constraint and its
charge sectors, and order by order in the lattice spacing through
Eq.~(\ref{eq:longwave}). It does not imply that the full real-time
evolution of mimetic gravity is reproduced by the fcc bosonic
Hamiltonian: demonstrating full dynamical equivalence would
require matching the complete Hamiltonian, the closure of the
constraint algebra, and the evolution equations of both theories,
none of which is attempted here. The present construction is
therefore a canonical embedding, not a full discretization of
mimetic gravity. The constraint sector is nevertheless the natural
invariant to match first: the constraints generate the gauge
symmetry of both theories, and because the effective lattice
Hamiltonian (\ref{ELEVEN}) is invariant under the rank-two gauge
transformation, the sourced Gauss law with a static defect is
preserved under the lattice time evolution, so the correspondence
is stable under the dynamics that the lattice model does possess.
This division of labor reflects the origin of the program. The
construction of Refs.~\cite{Gu:2009jh,Xu:2006faa,Xu:2010eg}, which
we extend, is quasi-static: it identifies the emergent spin-2
structure at low frequencies and large wavelengths, and finding
the lattice counterpart of the full gravitational dynamics is a
hard open problem already for the pure gravity sector. The
contribution of the present paper is therefore deliberately placed
at the level of the constraints, which a quasi-static construction
can capture exactly; the real-time dynamics of mimetic gravity and
of its dark sector on the lattice remains the main open problem of
this program.

A natural concern about any lattice realization of mimetic gravity
is whether the underlying continuum theory it discretizes is
itself physically well defined. The minimal mimetic dark-matter
model and several of its simplest higher-derivative extensions are
known to suffer from ghost and gradient instabilities in
cosmological perturbations. Specifically, Ijjas, Ripley and
Steinhardt \cite{Ijjas:2016pad} showed that mimetic cosmology is
generically prone to gradient instabilities even in regimes that
satisfy the null-energy condition, except for trivial examples,
and that matter stress-energy does not by itself supply the
stabilising gradient structure that is missing in the bouncing
constructions of \cite{Chamseddine:2014vna,Chamseddine:2016uef}.
Firouzjahi, Gorji and Mansoori \cite{Firouzjahi:2017txv} then
established that, in a broad class of mimetic extensions with
higher-derivative terms, scalar perturbations carry wrong-sign
kinetic or gradient terms; the resulting pathology is not an
ordinary Ostrogradsky ghost. Zheng \emph{et al.}
\cite{Zheng:2017qfs} reached the compatible perturbative
conclusion that simple higher-derivative extensions cannot cure
the instability on their own, though direct couplings of higher
mimetic derivatives to spacetime curvature can evade it; a
related Hamiltonian analysis was given subsequently in
\cite{Zheng:2018cuc}. In the same direction, Takahashi and
Kobayashi \cite{Takahashi:2017zoq} showed that extended mimetic
theories are typically plagued by either tensor/scalar gradient
instabilities (except in a likely strongly-coupled corner) or by
ghost instabilities, providing a complementary no-go diagnostic.
Against this background, the first explicit healthy reformulation
was proposed by Hirano, Nishi and Kobayashi \cite{Hirano:2017zox},
who showed that an effective-field-theory construction in the
unitary gauge $\phi = t$ yields a healthy sound speed
$c_s^2 \geq 0$ provided the appropriate higher-order operators
are turned on.

The post-2019 literature has continued this program along
several converging routes. Langlois, Mancarella, Noui and Vernizzi
\cite{Langlois:2018jdg} showed that mimetic gravity, including the
higher-derivative completions above, can be viewed as a class of
degenerate higher-order scalar-tensor (DHOST) theories with an
extra local gauge symmetry, providing a systematic framework in
which the propagating degrees of freedom and the linear-perturbation
structure can be analyzed in EFT form. Ganz, Bartolo and Matarrese
\cite{Ganz:2019pkl} formulated a viable effective field theory of
mimetic gravity and pointed out a subtle but important caveat:
restricting the analysis to homogeneous scalar backgrounds can
disguise the number of propagating degrees of freedom present
generically, since an extra mode can re-appear for inhomogeneous
field profiles. From a strictly canonical viewpoint, de Cesare
and Husain \cite{deCesare:2020swb} reformulated mimetic gravity in
the dust-time gauge $\phi = t$ and made explicit the physical
Hamiltonian generating the evolution of the spatial geometry, in a
form directly compatible with the analysis of
Sec.~\ref{sec:prelim}. More recently, Dom\`enech and Ganz
\cite{Domenech:2023ryc,Domenech:2025bqe} have placed mimetic
gravity within the broader landscape of disformal-symmetric
theories and clarified its precise relationship with
unit-timelike-vector relativistic-MOND constructions such as
TeVeS and AeST. In the same period, simple mimetic scalar-field
dark-matter models that are simultaneously ghost-free and
caustic-free have been constructed explicitly
\cite{Kanambaye:2025srm}.
Independently, the multi-field mimetic literature has clarified
the degree-of-freedom counting and stability structure of
two-field and curved-field-space generalizations: Shen, Zheng and
Li \cite{Shen:2019nyp} showed that two-field mimetic gravity
carries two scalar degrees of freedom in addition to the tensor
modes and identified an opposite-sign-constraint sector in which
ghost instabilities appear; Hosseini Mansoori \emph{et al.}
\cite{Mansoori:2021bjj} extended the construction to multi-field
models on curved field-space and showed that the health of the
entropy perturbation depends on the field-space metric signature;
and Zheng and Rao \cite{Zheng:2022rad} broadened the two-field
mimetic constraint structure and found that the generalized setup
still mimics dark matter at the background level. Post-GW170817
compatibility of specific mimetic realizations of dark matter and
dark energy has also been analyzed in detail
\cite{Casalino:2018tcd,Casalino:2018wnc,GBM:2017lvd,Sakstein:2017xjx}.

Against this background it is important to distinguish three
logically distinct stability questions: (i) the constraint-level
map developed in this paper; (ii) the perturbative stability of
the continuum mimetic completion whose source is being mapped;
and (iii) the stability of the microscopic lattice Hamiltonian
that realizes the rank-two gauge structure. This paper addresses
(i). It does not claim that (ii) or (iii) are solved for the
minimal mimetic model. Concerning (iii), the fcc bosonic model
possesses its own low-energy stability within the algebraic
Bose-liquid phase, protected by the rank-two gauge symmetry
\cite{Xu:2006faa,Xu:2010eg}; but this does not amount to a
stability proof for a mimetic cosmological scalar sector.

With this decomposition, the question of whether the lattice
construction avoids, modifies, or inherits the continuum
instabilities admits a sharp answer. It \emph{inherits} them
in the naive continuum limit: the map is agnostic about the
completion, so if the minimal mimetic theory is used as input, its
ghost and gradient instabilities reappear at long wavelength. It
\emph{modifies} them only in the ultraviolet: on the lattice the
growth rates would be bounded: in a local real-time
discretization of the same unstable dispersion, the compact
Brillouin zone imposes a momentum cutoff $k \lesssim \pi/a$, which
regulates the ultraviolet growth but does not cure the
instability. No claim of ghost or gradient stability is made
without a perturbation analysis. And it \emph{avoids} them only in
the conditional sense that the same constraint-level strategy can
be reapplied after deriving the full momentum source of the
selected healthy completion: we
have established a continuum-to-lattice mapping for the canonical
source structure, in the scalar sector
$J^j_{\rm mim} = (1/2) p_\phi h^{ji}\partial_i\phi$, and in the
vector sector the analogous source of Eq.~(\ref{THIRTYTWO}), both
realized as defects of the underlying Gauss law. This statement
depends primarily on the spatial-diffeomorphism momentum
constraint, which is robust under any of the formulations cited
above. In the minimal model the canonical momentum takes the
simple form $p_\phi=\sqrt{h}\,\lambda\,\nabla_n\phi$ and the
lattice dictionary applies directly; in higher-derivative or DHOST
completions the map remains valid with $p_\phi$ interpreted as the
full canonical momentum of the chosen completion. The lattice map
should therefore not be read as a proof that the minimal continuum
mimetic theory is itself a viable cosmological model. Rather, it
provides a kinematic dictionary that can be carried over to
whichever mimetic completion is selected on stability grounds.

Two consequences of this scoping deserve emphasis. First, the
lattice defect studied here is intrinsically inhomogeneous, so the
caveat raised in \cite{Ganz:2019pkl} about analyses restricted to
homogeneous backgrounds is directly relevant: stability arguments
derived only around FLRW backgrounds may need to be revisited in
the presence of the inhomogeneous source. Second, a complete
cosmological treatment of any of the models in
Sec.~\ref{sec:grat} would require a perturbation analysis around
the chosen healthy completion, with explicit checks of the scalar
sound speed $c_s^2$ and of the tensor speed $c_T^2 = 1$ in the
linearized sector. Such an analysis is outside the scope of the
present formal map and is left for follow-up work, for which the
dust-time canonical framework of \cite{deCesare:2020swb} and the
DHOST formulation of \cite{Langlois:2018jdg} are the natural
starting points.

\subsection{Cosmological outlook}
\label{subsec:cosmoout}

Although the present construction is quasi-static and canonical,
it is useful to indicate how its ingredients would enter a
cosmological application, and to delimit what such an application
would require. In the dust-time gauge $\phi = t$
\cite{deCesare:2020swb} the spatial gradients of the mimetic
scalar vanish and the homogeneous sector contributes through the
Hamiltonian constraint: this is the sector in which the continuum
theory reproduces the known mimetic dust cosmology
\cite{Chamseddine:2013kea,Chamseddine:2014vna,Dutta:2017fjw}, with
dark-matter-like background behavior, and in which
GW170817-compatible mimetic models of the combined dark sector
have been constructed \cite{Casalino:2018tcd,Casalino:2018wnc}.
The defect sector studied in this paper is the complementary,
intrinsically inhomogeneous one: a coarse-grained density of
lattice defects corresponds, through
Eqs.~(\ref{NINETEEN}) and (\ref{eq:longwave}), to a distribution
of dust-like momentum sources, and the two-field regimes of
Sec.~\ref{subsec:inhodarkener}, labeled by the ratio
$p_\phi/p_\psi = 2\lambda/\omega$ (valid, as established there, in
the spatial-gradient-orthogonal or cross-gradient-subleading
regime), would translate into a
qualitative interpolation between dark-matter-dominated and
dark-energy-dominated source content. We emphasize that this
paragraph is a dictionary statement, not a cosmological solution.

This program should be contrasted with cosmological model-building
studies of the dark sector in modified gravity, where one studies
background evolution, effective equations of state,
thermodynamics, or observational diagnostics: mimetic $f(R)$ and
$f(R,T)$ cosmologies \cite{Momeni:2015mimetic}, anisotropic
mimetic cosmologies \cite{GudekliDemir:2021}, and holographic
dark-energy scenarios built on modified entropy--area laws of the
Tsallis and Barrow type
\cite{ZubairMuneerGudekli:2022,SultanaEtAl:2024,ChakrabortyEtAl:2021}.
The present paper addresses a different and complementary
problem, the canonical lattice realization of mimetic source
terms, but the comparison suggests one genuinely microscopic question
raised by the holographic scenarios: modified entropy--area laws
are commonly motivated by quantum-gravitational deformations of
horizon entropy, and qubit lattice models are precisely the
setting in which entanglement entropy and its corrections to the
area law can be computed from first principles. Whether the
entanglement entropy of the Gauss-law defect sectors constructed
here exhibits such deformations is an open question that we leave
for future work.

A complete cosmological treatment based on the present
construction would require, at a minimum: real-time lattice
dynamics beyond the quasi-static regime; a coarse-graining
prescription connecting defect densities to homogeneous ADM data;
an FLRW reduction with the resulting background equations and an
effective equation of state; and a perturbation analysis around a
healthy completion, with the stability checks of
Sec.~\ref{sec:stability-scope}. None of these steps is attempted
here.


\section{Final Remarks}
\label{sec:final}

In the present article we have constructed a canonical lattice
realization of mimetic dark-sector sources, emergent in the
constraint-level sense defined in Sec.~\ref{subsec:physcontent}, using
the fcc bosonic qubit model of Refs.~\cite{Xu:2006faa,Xu:2010eg}
(originally introduced to study emergent gravitons and
Horava-Lifshitz gravity) as the gravitational substrate. We close by stating explicitly
what has been achieved, what the limitations of the construction
are, and which extensions it suggests.

\medskip
\noindent {\it Achievements.} First, the mimetic constraint was
treated in its full ADM form, Eq.~(\ref{eq:ADMmimetic}), retaining
the normal derivative of the scalar field, with the Lagrange
multiplier handled canonically on the constraint surface; this is
the formulation suitable for discretization. Second, the canonical
source terms of four mimetic sectors, namely scalar
\cite{Chamseddine:2013kea}, two-scalar
(Sec.~\ref{subsec:inhodarkener}), vector
(Sec.~\ref{subsec:vectormimedarkmatt}), and tensor-vector-scalar
(Sec.~\ref{subsec:mimeteves}), were mapped onto sources of the
rank-two lattice Gauss law, Eqs.~(\ref{NINETEEN}),
(\ref{TWENTYFOUR}), (\ref{THIRTYTWO}) and (\ref{THIRTYSIX}), with
the continuum-lattice dictionary summarized in
Table~\ref{tab:dictionary} and verified in the long-wavelength
limit, Eq.~(\ref{eq:longwave}). Third, the identification has
physical content beyond the dictionary: the lattice Gauss law
imposes the selection rules
(\ref{eq:chargerule})--(\ref{eq:momentrule}), which classify the
admissible defect configurations on a closed lattice, and the
worked example of Sec.~\ref{subsec:pointdefect} shows that a
localized defect sources a long-range $1/r^2$ rank-two field
whose trace-free representative is exactly the Bowen--York
momentum solution, after matching the source normalization: at
the constraint level, the defect sources nontrivial gravitational
canonical momentum.
The reduced-phase-space regimes of the two-scalar sector, labeled
by the multiplier ratio $2\lambda/\omega$, organize the relative
weight of dark-matter-like and dark-energy-like source content.

\medskip
\noindent {\it Limitations.} The correspondence established here
is a canonical embedding, not a full discretization of mimetic
gravity: no equivalence of the real-time dynamics, of the total
Hamiltonian, or of the constraint algebra beyond the momentum
sector is claimed. The construction is quasi-static and does not
provide a cosmological background evolution, an equation of state,
or observational constraints. The lattice regulates but does not
cure the ghost and gradient instabilities of the minimal continuum
mimetic theory (Sec.~\ref{sec:stability-scope}); health must be
supplied by the choice of continuum completion, and the
constraint-level strategy is then reapplied to the full momentum
source of that completion. Finally, individual lattice defects are the
canonical representatives of mimetic momentum-density sources; they
are not identified with phenomenological dark-matter particles or
dark-energy fluid elements.

\medskip
\noindent {\it Future directions.} The main open problem of this
program is the construction of the real-time lattice dynamics
beyond the quasi-static regime, which is nontrivial already for
the pure spin-2 sector of
Refs.~\cite{Gu:2009jh,Xu:2006faa,Xu:2010eg}. The further natural
continuations are: carrying the map to a healthy mimetic completion (DHOST,
effective-theory, or disformal
\cite{Hirano:2017zox,Langlois:2018jdg,Ganz:2019pkl,deCesare:2020swb}),
including the higher-derivative theories of Horava-Lifshitz type
\cite{Cognola:2016gjy,Chamseddine:2019gjh}; the cosmological
reduction and perturbation analysis outlined in
Sec.~\ref{subsec:cosmoout}; and the study of whether the restricted
mobility suggested by the selection rules
(\ref{eq:chargerule})--(\ref{eq:momentrule}) produces distinctive
dark-sector signatures, together with the entanglement entropy of
the defect sectors, where lattice models offer first-principles
access \cite{Pretko:2017fbf,You:2019cvs}. This program is part of
a broader project within the lattice framework for field theories
developed mainly by Wen \cite{Wen:2017usd}.

Two recent developments in particular suggest that the kinematic
bridge developed here can be exploited further. On the continuum
side, Dom\`enech and Ganz \cite{Domenech:2025bqe} have shown that
relativistic-MOND theories featuring a unit-timelike vector, such
as TeVeS and AeST, can be embedded in a conformal-disformal
framework closely related to mimetic gravity, so that the present
lattice map is naturally extendable to the
relativistic-MOND-survivor framework provided by AeST
\cite{Skordis:2020eui,Skordis:2021ahj}. On the condensed-matter
side, the rank-two fcc construction belongs to the broader family
of higher-rank gauge theories whose connection with fracton
physics has been clarified in recent reviews
\cite{Pretko:2017xar,You:2024prv}; an explicit fractonic
realization of emergent matter fields and mimetic gravity
compatible with our construction has appeared in
\cite{Wang:2019aiq}. Combining these threads is a natural avenue
for follow-up work.

 \vspace{.5cm}
\centerline{\bf Acknowledgments} \vspace{.5cm}

L. L. would like to thank CONACyT for a grant with CVU number
594425. It is a pleasure to thank Norma Quiroz for useful
conversations.


\end{document}